\documentclass[twocolumn,superscriptaddress,aps,preprintnumbers,amsmath,amssymb,prl,nofootinbib]{revtex4}

\usepackage{mathrsfs,amsmath}
\usepackage{mathtools}
\usepackage{empheq}
\usepackage[pdftex]{graphicx}
\usepackage[dvipsnames]{xcolor}
\usepackage{float}
\usepackage{physics}
\usepackage{comment}
\usepackage{braket}
\usepackage[normalem]{ulem}
\usepackage{caption}
\usepackage{subcaption}
\usepackage{cancel}
\usepackage[dvipsnames]{xcolor}
\usepackage[normalem]{ulem}

\usepackage[colorlinks,linkcolor=blue,anchorcolor=violet,citecolor=red]{hyperref}

\usepackage{bm}

\newcommand*{\tU}{{\widetilde{\mathcal U}}}
\newcommand{\ini}{{\rm ini}}
\newcommand{\pos}{{\rm post}}
\begin{document}

\title{
Effective Time Advance of Quantum Clocks under Postselection
}

\author{Youka Kaku}
\email{ykaku@stevens.edu}
\affiliation{
Department of Physics, Stevens Institute of Technology, Hoboken, New Jersey 07030, USA}
\affiliation{
Research Center for the Early Universe (RESCEU), Graduate School of Science, University of Tokyo, Bunkyo, Tokyo 113-0033, Japan}
\author{Akira Matsumura}
\email{matsumura.akira@phys.kyushu-u.ac.jp}
\affiliation{Department of Physics, Kyushu University, Fukuoka 819-0395, Japan}
\affiliation{Quantum and Spacetime Research Institute, Kyushu University, Fukuoka 819-0395, Japan}
\author{Tomohiro Fujita}
\email{fujita.tomohiro@ocha.ac.jp}
\affiliation{Department of Physics, Ochanomizu University, Bunkyo, Tokyo 112-8610, Japan}
\affiliation{Kavli Institute for the Physics and Mathematics of the Universe (Kavli IPMU),
WPI, UTIAS, The University of Tokyo, Kashiwa, Chiba 277-8568, Japan}

\date{\today}

\begin{abstract}
In the familiar twin paradox, the traveling twin ages less than the one who stays behind as a consequence of relativistic time dilation. Here, we show that quantum postselection can effectively reverse this outcome. Using an analytically solvable quantum clock model, we demonstrate that postselection of the center-of-mass motion produces an effective time advance through anomalous weak values, even though every individual motional branch undergoes ordinary time dilation. We extend this analysis to a weak gravitational field and show that an effective time advance can arise even for simple coherent states of the quantum clock motion. We further illustrate this effect using a hydrogen-like atom, revealing how it manifests itself in the atomic wavefunction. Our results demonstrate that quantum interference and postselection can produce an effective time advance beyond conventional relativistic time dilation.
\end{abstract}

\maketitle

\section{I. Introduction}


The interplay between quantum mechanics and relativity gives rise to fundamental questions about nature, ranging from how to formulate a quantum theory of gravity to how quantum probes behave in curved spacetime.
One of the most striking lessons of relativity is that time does not flow at a universal rate. 
In special relativity, the twin paradox illustrates how a traveling clock accumulates less proper time than a clock that remains at rest. In general relativity, a clock at a lower gravitational potential ticks more slowly than one at a higher potential. Both effects have been confirmed by atomic-clock experiments \cite{Hafele1972,Takamoto2020,Bothwell2022} and are essential to the operation of the Global Positioning System (GPS) \cite{Ashby2003}. This raises a natural question: how does relativistic time dilation manifest when the clock itself is treated as a quantum system?

To address this question, we consider quantum clocks, which have been studied as composite quantum systems consisting of center-of-mass and internal degrees of freedom \cite{zych2011quantum,zych2019gravitational}. 
For example, an atom can serve as a quantum clock \cite{marshall2025high,Arnold2026}; oscillations between different internal energy states provide a precise measure of proper time, including relativistic effects.
In particular, when the internal state is prepared in a superposition of different energy levels, mass-energy equivalence implies that the clock is in a superposition of different rest masses. 
Consequently, the center-of-mass motion evolves under different redshift factors depending on the internal energy level. As a result, the center-of-mass and internal degrees of freedom become entangled, a phenomenon with no counterpart in classical theories.
Previous studies have shown that gravitational time dilation can reduce interferometric visibility and induce decoherence by tracing out the center-of-mass motion\cite{zych2011quantum,haustein2019mass,Pikovski2015,kaku2022quantumness}. Related effects arising from special relativistic time dilation have also been investigated for trapped-ion clocks, including their potential experimental observation \cite{Sorci2026}.

Another intriguing phenomenon arising from the interplay between gravity and quantum mechanics is the gravitational repulsive force through postselection \cite{bose2026effective,saldanha2026repulsive}. Although the gravitational force is attractive in each branch of a quantum superposition, their superposition can effectively produce a repulsive force under appropriate postselection. This counterintuitive behavior arises from quantum interference and can be understood in terms of anomalous weak values, which can lie outside the eigenvalue range of an observable \cite{Aharonov1988,Dressel2014}.
Inspired by the idea of effective gravitational repulsion in quantum mechanics, one may expect similar counterintuitive behavior in relativistic phenomena, such as gravitational redshift and time dilation. The basic concept of anomalous time translations through postselection was explored in earlier studies \cite{Aharonov1990,Vaidman1991}, which showed that a postselected quantum state can effectively evolve as if time had passed differently, in ways that cannot be realized classically.


Motivated by these studies, we investigate how an anomalous passage of time can emerge in the relativistic dynamics of quantum clocks. We show that anomalous time evolution can arise from both special relativistic and gravitational redshift effects through postselection. In classical relativity, the redshift factor is given by $1-p^2/(2m^2c^2)+\Phi/c^2$ up to order $1/c^2$, where the second and third terms represent the special relativistic and gravitational contributions with the weak gravitational potential $\Phi(<0)$, respectively. Thus, the redshift factor can never exceed unity in this classical setting. However, when postselection is performed on a quantum clock, the effective shift factor can exceed unity under certain conditions, leading to an effective time advance rather than the usual time delay.

We first focus on special relativistic time delay and advance in the absence of a gravitational potential. We consider a setup inspired by the twin paradox, in which each twin carries a quantum clock with two internal energy levels to measure their proper time. We then introduce a Ramsey interferometry protocol \cite{ramsey1950molecular,ludlow2015optical} to explain how each twin can extract their proper time from the quantum clock state.
In the ordinary procedure without postselection, the proper time of the traveling twin is given by $t\,\Big(1-\langle \hat p^2\rangle/(2m^2c^2)\Big)$, which cannot exceed the reference proper time $t$ of the stationary twin, resulting in the usual time delay. On the other hand, when postselection is performed during the Ramsey interferometry protocol, the effective proper time of the traveling twin is given by $t\,\Big(1-\text{Re}[(\hat p^2)_w]/(2m^2c^2)\Big)$, where $(\cdots)_w$ denotes the weak value. Since the weak value can be any complex number in general, the effective proper time can exceed the reference time $t$ of the stationary twin, leading to an effective time advance in contrast to the ordinary twin paradox.
We then extend the analysis to gravitational time delay and advance. We also show that effective time advance can occur even when the quantum clock is initially prepared in a coherent state and postselected onto another coherent state, without requiring an extreme superposition state.
Finally, we discuss the physical picture of time advance using a hydrogen-like atom as a quantum clock and show how this phenomenon can be understood in terms of the wavefunction dynamics.


This paper is organized as follows. In Sec. II, we review special relativistic time dilation for a quantum clock using the twin paradox. In Sec. III, we show that postselection of the center-of-mass state can lead to an effective time advance. In Sec. IV, we extend the analysis to a weak gravitational field. In Sec. V, we investigate time advance using coherent states, for which the time evolution can be solved analytically. In Sec. VI, we provide a physical picture of time advance by examining the internal dynamics of a hydrogen-like atom. Finally, we summarize our results in Sec. VII.

\section{II. Special relativistic time delay of the traveler in the twin paradox}

In this section, we revisit the twin paradox using quantum clocks, whose center-of-mass motion and internal dynamics are both treated quantum mechanically. We consider two twins, Alice and Bob, who carry quantum clocks and compare their elapsed proper times through Ramsey interferometry. We show that the traveling twin Bob experiences the usual relativistic time dilation, even when his clock is described quantum mechanically. This analysis provides the foundation for our discussion in the subsequent sections.

\subsection{Quantum clocks}

We consider a quantum clock in special relativity. 
A quantum clock has center-of-mass (COM) and internal degrees of freedom.
Here, we focus on a one-dimensional COM motion for simplicity.
In the nonrelativistic regime, the clock Hamiltonian is given by
\begin{align}
    \hat H=
    \frac{\hat p^2}{2m}
    +\hat H_\text{int} \Big(1-\frac{\hat{p}^2}{2m^2 c^2} \Big),
    \label{eq:H}
\end{align}
where 
$\hat p$ is the COM momentum of the clock, and 
\begin{align}
 \hat H_\text{int}= \hbar \omega |e\rangle\langle e|
    \label{eq:Hc}
\end{align}
is the internal Hamiltonian. 
The internal degree of freedom is described by a two-level system with a ground state $|g\rangle$ and an excited state $|e\rangle$
, whose 
energy eigenvalues are $0$ and $\hbar \omega$, respectively. 
The Hamiltonian $\hat H$ in Eq.~\eqref{eq:H} is 
derived from 
mass--energy equivalence. 
Starting from the free Hamiltonian $mc^2+\hat{p}^2/(2m)$, we replace the mass $m$ with the total rest mass including the internal energy $m+\hat{H}_\text{int}/c^2$, expand in $1/c$, and then neglect the constant offset $mc^2$. This yields Eq.~\eqref{eq:H}. 

The quantum clocks carried by Alice and Bob are both described by the Hamiltonian in Eq.~\eqref{eq:H}.
The redshift factor $1-\hat p^2/(2m^2 c^2)$ couples the COM momentum $\hat p$ to the internal Hamiltonian $\hat H_\text{int}$. 
Consequently, the COM undergoes different time evolutions depending on whether the clock is in its ground state $|g\rangle$ or excited state $|e\rangle$. 
This coupling can generate entanglement between the COM and internal degrees of freedom.


\subsection{Proper time measurement using Ramsey interferometry}

\begin{figure*}[htbp]
    \centering
    \includegraphics[width=0.9\linewidth]{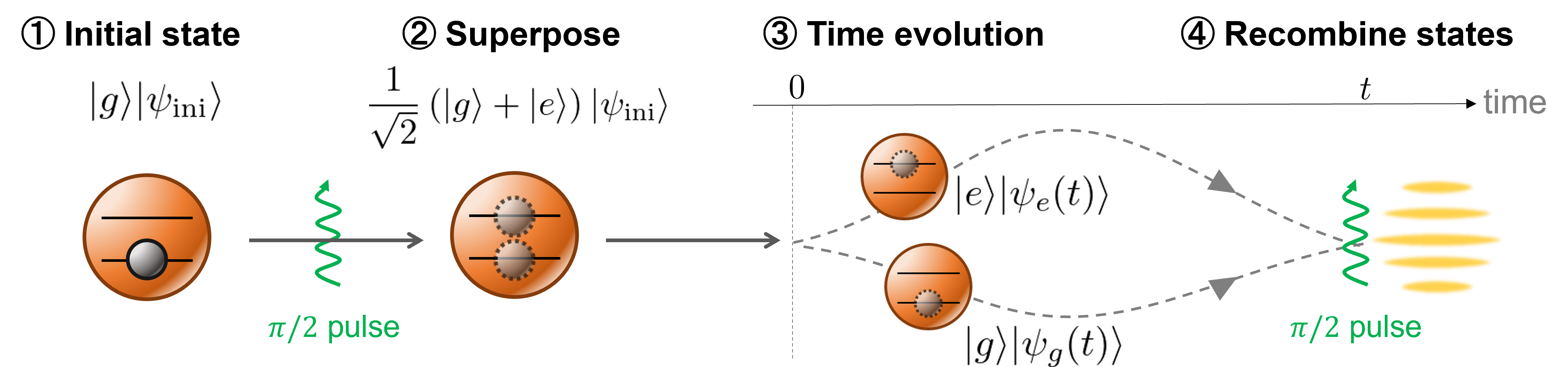}
    \caption{
    Four-step Ramsey interferometry protocol for a quantum clock. (1) Prepare the clock in its internal ground state. (2) Apply a $\pi/2$ pulse to create a superposition of the ground and excited states. (3) Let the clock evolve for a time $t$. (4) Apply a second $\pi/2$ pulse and measure the excitation probability to extract the proper time $\tau$ in Eq.~\eqref{eq:tau_aprx}.}
    \label{fig:Ramsey}
\end{figure*}

Here, we explain how to extract the proper time from a quantum clock. For this purpose, we use Ramsey interferometry, which allows us to probe the internal state of the clock particle. As shown in Fig.~\ref{fig:Ramsey}, the Ramsey protocol consists of four steps.

First, we prepare the initial state
\begin{align}
    |\Psi_\ini\rangle=
    |g\rangle
    |\psi_\ini\rangle,
\end{align}
where $|g\rangle$ denotes the ground state of the internal degree of freedom and $|\psi_\text{ini}\rangle$ is the initial COM state.

Second, we apply a $\pi/2$ pulse to prepare a superposition of the two internal energy states:
\begin{align}
    |\Psi'_\ini\rangle
    =\hat R \, |\Psi_{\rm ini}\rangle
    =\frac{1}{\sqrt{2}}
    \left(|g\rangle+|e\rangle\right)
    |\psi_{\rm ini}\rangle
    \label{eq:Ramsey2}
\end{align}
Here, $\hat R=\exp\left[-i\hat \sigma_y \pi/4\right]$ is the pulse operator, which rotates the state by $\pi/2$ on the Bloch sphere. This rotation can be achieved using an external electric field for atomic electronic states or a magnetic pulse for spin states.

Third, we let the state $|\Psi'_\ini\rangle$ evolve for a time $t$ under the Hamiltonian \eqref{eq:H}:
\begin{align}
    |\Psi(t)\rangle
    :=\hat U(t) |\Psi'_\ini\rangle
    =\frac{1}{\sqrt{2}}
    \left(|g\rangle|\psi_g(t)\rangle+|e\rangle|\psi_e(t)\rangle
    \right)
    ~,
    \label{eq:Ramsey3}
\end{align}
where $\hat U(t)=e^{-i\hat H t/\hbar}$ is the time evolution operator, and \begin{align}
  |\psi_g\rangle&=\exp\Big[-i \frac{\hat{p}^2t}{2m\hbar}\Big]|\psi_\text{ini}\rangle,
   \label{eq:psi_g}
   \\
   |\psi_e\rangle&=\exp\Big[-i \frac{\hat{p}^2t}{2m\hbar}-i\omega t\Big(1-\frac{\hat{p}^2}{2m^2 c^2} \Big)\Big]|\psi_\text{ini}\rangle
   \label{eq:psi_e}
\end{align} 
denote the COM states associated with the internal energy levels $|g\rangle$ and $|e\rangle$, respectively. 
The COM and the internal degrees of freedom are thus entangled. 

Fourth, we apply a second $\pi/2$ pulse to recombine the internal states:
\begin{align}
    |\Psi'(t)\rangle
    &=\hat R\, |\Psi(t)\rangle
    \notag\\
    &=
    \frac{|g\rangle+|e\rangle}{2}\,|\psi_g(t)\rangle
    +
    \frac{-|g\rangle+|e\rangle}{2}\,|\psi_e(t)\rangle.
    \label{eq:Ramsey4}
\end{align}
We then measure the probability of finding the clock in the excited state $|e\rangle$:
\begin{align}
    \mathcal{P}=\big\|\langle e|\Psi'(t)\rangle\big\|^2
    =\frac{1}{2}\Big(
    1+\mathrm{Re}\big[
    \langle \psi_g (t)|\psi_e (t)\rangle
    \big]\Big).
    \label{eq:Ramsey_prob}
\end{align}
The overlap $\langle \psi_g(t)|\psi_e(t)\rangle$ is called the Ramsey interference term. 
Since these states evolve under different Hamiltonians, their overlap contains information about the internal energy gap $\hbar \omega$ and the redshift factor $1-\hat p^2/(2m^2 c^2)$.

We now show how to extract the proper time from Ramsey interferometry. 
Assuming that the internal transition frequency $\omega$ is known or independently calibrated, the proper time can be estimated from the phase of the Ramsey interference term as
\begin{align}
    \tau
    &=-\frac{1}{\omega} \arg\langle \psi_g|\psi_e\rangle
    \simeq
     t\left(
    1
    -
    \frac{\langle \psi_\text{ini}|
    \hat{p}^2 
    |\psi_\text{ini}\rangle}{2m^2 c^2}
    \right)
    \label{eq:tau_aprx}
    ~.
\end{align}
In the second equality, we expanded it to the first order in $\hat{p}^2 \omega t/(2m^2 c^2) \ll 1$ 
and used the fact that $\langle \psi_\text{g}|\hat{p}^2 |\psi_\text{g}\rangle=\langle \psi_\text{ini}|\hat{p}^2 |\psi_\text{ini}\rangle$.

\subsection{Time delay of the traveler in the twin paradox}

We illustrate the twin paradox by comparing the proper times of Alice and Bob measured using Ramsey interferometry.
We assume that the initial COM state of Alice's clock is a wavepacket sharply peaked around zero momentum, with a sufficiently small momentum spread that its relativistic time-dilation effect can be neglected, $|\psi^A_\text{ini} \rangle\simeq |p=0\rangle$.
She therefore measures the proper time as follows:
\begin{align}
    \tau^A
    \simeq
    t~.
    \label{eq:tauA}
\end{align}
On the other hand, we assume that Bob travels with nonzero momentum and measures his proper time as follows:
\begin{align}
    \tau^B\simeq  t\left(
    1
    -
    \frac{\langle \psi^B_\text{ini}|\hat{p}^2|\psi^B_\text{ini}\rangle}{2m^2 c^2}
    \right)
    ~,
    \label{eq:tauB}
\end{align}
where $|\psi^B_\text{ini}\rangle$ is the COM state of Bob's clock, which has distribution around the nonzero momentum. 
Since $\hat p^2$ is a nonnegative operator, its expectation value 
$\langle \psi^B_\text{ini}|\hat{p}^2|\psi^B_\text{ini}\rangle$ is also nonnegative. We therefore obtain
\begin{align}
    \tau^B \leq \tau^A
    ~.
    \label{eq:timedelay}
\end{align}
Thus, the proper time of Bob's clock is shorter than or equal to that of Alice's clock. 
Given that the twins' ages are operationally defined by the elapsed times of their quantum clocks,
this means that the traveling twin is younger than or the same age as the stationary twin, consistent with the usual twin paradox.


\section{III. Effective special relativistic time advance under postselection}

In this section, we show that postselection of Bob's quantum clock can yield an effective time advance. This indicates that the traveling twin Bob can be older than the stationary twin Alice in contrast to the ordinary twin paradox.

Let us suppose that the traveler Bob performs a postselection onto the COM state of his quantum clock, where the postselected state is denoted as $|\psi_\text{fin}^B\rangle$. This postselection is performed just after the third step of the Ramsey interferometry, that is on the time evolved state $|\Psi^B(t)\rangle$ given in Eq.~\eqref{eq:Ramsey3}. Then, Bob's internal state after the postselection is given by
\begin{align}
    \ket{\chi_\pos^B}
    :=
    \frac{\langle\psi_\text{fin}^B|\Psi^B(t)\rangle}{\left\|\langle\psi_\text{fin}^B|\Psi^B(t)\rangle\right\|}
    =
    N
    \left[
        |g\rangle
        +
        \frac{\langle\psi_\text{fin}^B|\psi_e^B\rangle}{\langle\psi_\text{fin}^B|\psi_g^B\rangle}        |e\rangle
    \right],
    \label{eq:Estate_PS}
\end{align}
where $N=\bigg(1+\bigg|\frac{\langle\psi_\text{fin}^B|\psi_e^B\rangle}{\langle\psi_\text{fin}^B|\psi_g^B\rangle}  \bigg|^2\bigg)^{-1/2}$ is a normalization factor. 
After applying the magnetic pulse $\hat R$ to recombine the superposition of the internal states $|\chi_\pos^B\rangle$, 
we obtain the excitation probability as 
\begin{align}
    \mathcal{P}_\pos^B
    =|\langle e |\hat R|\chi_\pos^B\rangle|^2
    =
    \frac{1}{2}\Big(1+2N^2
    \mathrm{Re}\left[
    \frac{\langle\psi_\text{fin}^B|\psi_e^B\rangle}{\langle\psi_\text{fin}^B|\psi_g^B\rangle} 
    \right] \Big)
    ~.
    \label{eq:Ramsey_prob_PS}
\end{align}
We can see that $2N^2\frac{\langle\psi_\text{fin}^B|\psi_e^B\rangle}{\langle\psi_\text{fin}^B|\psi_g^B\rangle}$ corresponds to the Ramsey interference term for the present case.
The effective proper time read off by Bob is now given by
\begin{align}
    \tau_\pos^B
    &=-\frac{1}{\omega} \arg\frac{\langle\psi_\text{fin}^B|\psi_e^B\rangle}{\langle\psi_\text{fin}^B|\psi_g^B\rangle}
    \simeq t\,
    \left(
    1
    -
    \frac{\text{Re}[(\hat{p}^2)_w]}{2m^2 c^2}
    \right)
    ~,
    \label{eq:tau_pos}
\end{align}
where we defined the weak value
\begin{align}
    (\cdots)_{w}
    :=
    \frac{
    \bra{\psi_\text{fin}^B}\cdots\ket{\psi_{g}^B}
    }
    {
    \braket{\psi_\text{fin}^B|\psi_g^B}
    }
    ~.
    \label{eq:weakvalue}
\end{align}
Most importantly, weak value generally gives a complex number in contrast to the expectation value. 
Therefore, although $\hat p^2$ gives nonnegative values in the usual Ramsey interference, the real part 
$\mathrm{Re}\left[(\hat p^2)_w\right]$ in Eq.~\eqref{eq:tau_pos} can now become negative after the postselection.
The negative $\mathrm{Re}\left[(\hat p^2)_w\right]$ typically arises from 
the interference between the initial and final states, $\ket{\psi_{g}^B}$ and $\ket{\psi_\text{fin}^B}$. 

Let $|0\rangle$ and $|p\rangle$ denote normalized COM wavepacket states sharply peaked around momenta $0$ and $p$, respectively, such that $\hat{p}|0 \rangle \simeq 0$ and $\hat{p}|p\rangle \simeq p|p\rangle$. 
These are assumed to be orthogonal to each other, $\langle 0|p\rangle \simeq 0$.  
We choose $|\psi^B_\text{ini} \rangle=\sqrt{\lambda}|0\rangle+\sqrt{1-\lambda}|p\rangle$ with $0\leq \lambda \leq 1$ and $|\psi^B_\text{fin}\rangle=(|0\rangle-|p\rangle)/\sqrt{2}$. 
Here, we focus on the timescale of the internal oscillation $t\sim 1/\omega$
and assume $p^2t/(2m\hbar)\ll1$, so that the free COM evolution can be neglected. 
Then, the weak value is 
\begin{align}
 (\hat{p}^2)_w \simeq -\frac{\sqrt{1-\lambda}}{\sqrt{\lambda}-\sqrt{1-\lambda}} \,p^2,
 \label{eq:weak_p2}
\end{align}
and the effective proper time $\tau^B_\pos$ is 
\begin{align}
 \tau^B_\pos=t \Big[1+\frac{\sqrt{1-\lambda}}{\sqrt{\lambda}-\sqrt{1-\lambda}}\frac{p^2}{2m^2 c^2} \Big].
 \label{eq:tau_rel}
\end{align}
When $\sqrt{\lambda} > \sqrt{1-\lambda}$, Eq.~\eqref{eq:tau_rel} gives
\begin{align}
    \tau_\pos^B > \tau^A
    ~,
\end{align}
where $\tau^A$ is the proper time measured by the stationary twin Alice as given in Eq.~\eqref{eq:tauA}.
Thus, Bob can be older than Alice even after his travel by performing the postselection in contrast to the time delay of the traveler in the ordinary twin paradox as given in Eq.~\eqref{eq:timedelay}.
We refer to this effect as an effective time advance, in contrast to the time delay.

So far, we have neglected the free COM evolution and used a leading-order approximation to the postselected phase for simplicity. However, these approximations are not essential for the effective time advance. First, the free COM evolution can be exactly compensated by choosing a time-dependent postselected state,
$|\psi^B_{\rm fin}(t)\rangle=(|0\rangle-e^{-i p^2t/(2m\hbar)}|p\rangle)/\sqrt{2}$.
For this choice, the ratio of the postselection amplitudes is given by $e^{-i\omega t}(\sqrt{\lambda}-\sqrt{1-\lambda}\,e^{i\delta})/(\sqrt{\lambda}-\sqrt{1-\lambda})$,
where $\delta=\omega t p^2/(2m^2c^2)$.
The approximation in Eq.~\eqref{eq:tau_rel} is valid when
$\sqrt{1-\lambda}|\delta|\ll|\sqrt{\lambda}-\sqrt{1-\lambda}|$.
Nevertheless, the exact phase of the above amplitude ratio still exhibits an effective time advance for $\lambda>1/2$ and $0<\delta<\pi$, even beyond this approximation. Thus, the effective time advance is not restricted to the simplifying assumptions used in this section.

The fact that quantum superposition and postselection can lead to anomalous time evolution was already suggested in Refs.\cite{Aharonov1990, Vaidman1991}. 
Here, we provided a concrete realization of anomalous time in terms of quantum clock and twin paradox.

\section{IV. Gravitational time delay and advance}

In this section, we discuss both the special relativistic and gravitational time delay/advance in the weak gravitational field. The metric of a weak gravitational field is
\begin{align}
    &ds^2=-\left(1+\frac{2\Phi}{c^2}\right)c^2dt^2+\left(1-\frac{2\Phi}{c^2}\right)d\bm{x}^2,\\
    &\Phi(\bm{x})=-\frac{GM}{|\bm{d}+\bm{x}|}
    ~.
\end{align}
Here, $M$ is the source mass, and $|\bm{d}+\bm{x}|$ is the distance between the clock's COM and the source mass.
For simplicity, we focus on one-dimensional motion. Then, the Hamiltonian of the quantum clock particle in a nonrelativistic regime is given by
\begin{align}
    \hat H
    =
    \frac{\hat p^2}{2m}+m\Phi(\hat x)
    +\hat H_\text{int}\,
    \left[1-\frac{\hat p^2}{2m^2c^2}+\frac{\Phi(\hat x)}{c^2} \right].
    \label{eq:hamiltonian_GR}
\end{align}
Here, $1-\hat p^2/(2m^2c^2)+\Phi(\hat x)/c^2$ is the redshift factor, 
which contains both the special relativistic and gravitational effects. 


\subsection{Gravitational time delay}

As before, the proper time can be measured from the phase of the Ramsey interference, $\langle \psi_g|\psi_e\rangle$, as follows.
\begin{align}
    &\tau
    =-\frac{\arg\langle\psi_g|\psi_e\rangle}{\omega} 
    \label{eq:tau_GR}
    \\
    &=-\frac{1}{\omega} \arg
    \left\langle
    \mathcal{T}\!
    \exp\!\left[
    -i\omega
    \!\int_0^t \!dt' \left(1-\frac{\hat p^2_I(t') }{2m^2c^2}+\frac{\hat{\Phi}_I (t') }{c^2} \right)
    \right]
    \right\rangle
    \nonumber
    \\
    &\simeq 
    \int_0^t dt' \left(1-\frac{\langle \hat p^2_I(t') \rangle }{2m^2c^2}
    +\frac{\langle \hat{\Phi}_I (t') \rangle}{c^2}\right).
    \label{eq:tau_GR_aprx}
\end{align}
Here, $\hat{\Phi}_I (t)=\Phi(\hat{x}_I (t))$, $\mathcal T$ denotes the time ordering, and $\langle\cdots \rangle$ represents the expectation value for $|\psi_\ini\rangle$. 
The subscript $I$ denotes the interaction picture, defined by $\hat A_I(t):=e^{i\hat H_gt/\hbar}\hat Ae^{-i\hat H_gt/\hbar}$, where $\hat{H}_g=\hat p^2/(2m)+m\Phi(\hat x)$. 
In the third line, we expanded the exponential to first order, assuming that the redshift effects are sufficiently small. 
The final expression shows that the proper time inferred from Ramsey interference is given by the time integral of the relativistic time-dilation factor, which includes both kinematic and gravitational contributions.

Now, let us consider the twin paradox in the gravitational field.
Assuming that Alice remains stationary and experiences no gravitational field, 
her proper time $\tau_A$ is still given by Eq.~\eqref{eq:tauA}. In contrast, if Bob moves with nonzero momentum and experiences the gravitational field, his proper time $\tau^B$ is given by Eq.~\eqref{eq:tau_GR_aprx}, where the expectation value is taken over Bob's initial state $|\psi_\ini^B\rangle$. Since both $\hat p^2$ and $-\Phi(\hat{x})$ are nonnegative operators, their expectation values are also nonnegative. 
We therefore obtain
\begin{align}
    \tau^B\leq \tau^A
    ~,
\end{align}
indicating that Bob becomes younger than Alice due to special relativistic and gravitational 
time dilation without postselection.

\subsection{Gravitational time advance}

As before, we suppose that Bob performs a postselection onto the COM state of his quantum clock immediately after the third step of Ramsey interferometry.
Bob's state and the excitation probability after the postselection is given by Eq.~\eqref{eq:Estate_PS} and \eqref{eq:Ramsey_prob_PS}, respectively. Then, we can read out the effective proper time from the Ramsey interference after the postselection, which yields
\begin{align}
    \tau_\pos^B
    &=-\frac{1}{\omega}\arg\frac{\langle\psi_\text{fin}^B|\psi_e^B\rangle}{\langle\psi_\text{fin}^B|\psi_g^B\rangle}
    \label{eq:tau_PS_GR}
    \\
    &\simeq 
    \int_0^t dt' \text{Re}\left[1-\frac{\left(\hat p^2_I(t') \right)_{\tilde{w}} }{2m^2c^2}
    +\frac{ \left(\hat \Phi_I (t')\right)_{\tilde{w}} }{c^2}\right]
    \label{eq:tau_PS_GR_aprox}
\end{align}
where we introduced the weak value as follows:
\begin{align}
    (\cdots)_{\widetilde w}
    =
    \frac{
    \bra{\widetilde\psi_\text{fin}^B} \cdots\ket{\psi_\ini ^B}
    }
    {
    \braket{\widetilde\psi_\text{fin}^B|\psi_\ini ^B}
    }
    ,\qquad
    |\widetilde \psi_\text{fin}^B\rangle= e^{i\hat{H}_g t/\hbar} |\psi_\text{fin}^B\rangle
    ~.
\end{align}
Since weak values generally give a complex number, $\mathrm{Re}\big[(\hat p_I^2)_{\widetilde w}\big]$ and $-\mathrm{Re}\big[(\Phi_I)_{\widetilde w}\big]$ can be negative, in contrast to the usual expectation values. Then Eq.~\eqref{eq:tau_PS_GR_aprox} therefore gives an effective time advance,
\begin{align}
    \tau_\pos^B>\tau^A
    \,,
\end{align}
when the time integral of $\mathrm{Re}\big[(\hat p_I^2)_{\widetilde w}\big]$ and $-\mathrm{Re}\big[(\Phi_I)_{\widetilde w}\big]$ is negative.
Thus, Bob can be older than Alice even in the presence of special-relativistic and gravitational time-dilation effects, which is in contrast to the ordinary twin paradox.

Let us demonstrate effective time advance induced by gravitational shifts. 
For simplicity, we assume that 
the COM motion of Bob's clock is negligible throughout its time evolution ($\hat p \simeq 0$), and that its initial state is in a superposition of two spatially localized states, 
$|\psi^B_\text{ini} \rangle=\sqrt{\lambda}|H\rangle +\sqrt{1-\lambda}|L\rangle$ with $0\leq \lambda \leq 1$. 
Here, $|H\rangle$ and $|L\rangle$ are normalized states approximately localized around the higher position $x_H$ and the lower position $x_L$, respectively: $\hat x|H \rangle \simeq x_H |H\rangle$, $\hat x|L\rangle \simeq x_L |L\rangle$. They are nearly orthogonal to each other $\langle L|H\rangle \simeq 0$.
This assumption allows us to focus on the phase evolution, while neglecting the dynamical evolution of Bob's COM states.
We then postselect the COM state onto $|\psi^B_\text{fin} \rangle=(|H\rangle -|L\rangle)/\sqrt{2}$. 
This postselection can be implemented by coherently recombining the two spatial wavepackets and selecting the appropriate output port before the final Ramsey measurement.

Furthermore, we focus on the time scale of the internal oscillation, $t\sim 1/\omega$, and assume that the COM phase evolution between the two heights is negligible by requiring $m|\Phi(x_H)-\Phi(x_L)| t/\hbar \ll 1$.
Then, we obtain $|\widetilde{\psi}^B_\text{fin} \rangle \simeq e^{im\Phi(x_H)t/\hbar}|\psi^B_\text{fin}\rangle$, 
and the weak value of the gravitational potential is given by 
\begin{align}
 (\Phi(\hat{x}))_w \simeq \frac{\sqrt{\lambda}\Phi(x_H)-\sqrt{1-\lambda} \Phi(x_L) }{\sqrt{\lambda}-\sqrt{1-\lambda}}, 
 \label{eq:weak_Phi}
\end{align}
and the proper time $\tau^B_\pos$ is 
\begin{align}
 \tau^B_\pos=t \Big[1+\frac{\sqrt{\lambda}\Phi(x_H)-\sqrt{1-\lambda} \Phi(x_L) }{c^2(\sqrt{\lambda}-\sqrt{1-\lambda})}\Big].
\end{align}
For $x_H > x_L$, the limit $\lambda \rightarrow 1/2 +0$ provides a simple limiting case in which effective time advance occurs, with $\tau_\pos^B > \tau^A$.

\section{V. Time delay and advance in coherent states}

In the previous sections, we demonstrated the basic mechanism of effective time advance using superpositions of distinct momentum or position states, while neglecting the dynamical evolution of the COM. Here, we go beyond these idealized setups and show that effective time advance can arise even when both the initial and postselected COM states are simple coherent states, corresponding to Gaussian wavepackets. This demonstrates that specially prepared Schrödinger-cat-like superpositions are not necessary for the effect.

To this end, we expand the gravitational potential up to quadratic order in the COM position, allowing us to solve the time evolution analytically. Although the Hamiltonian \eqref{eq:hamiltonian_GR} is derived within a leading-order relativistic approximation in $1/c^2$, its time evolution is treated exactly, without a perturbative Dyson expansion \eqref{eq:tau_GR_aprx}. This allows us to capture the accumulated effects of relativistic time dilation beyond the regime of a low-order expansion of the time evolution. Using this solution, we evaluate the Ramsey interference and the effective proper time after postselection.

First, we expand the gravitaitonal potential to second order in $x/d$:
\begin{align}
    \Phi(x)
    \simeq-\frac{GM}{d} \left(1-\frac{x}{d}+\frac{x^2}{d^2} \right).
    \label{eq:U_expansion}
\end{align}
The Hamiltonian can then be approximated by the following quadratic form:
\begin{align}
    \hat H
    &=
    \frac{\hat p^2}{2m}-\frac{GmM}{d} \left(1-\frac{x}{d}+\frac{x^2}{d^2} \right)
    \notag\\
    &+\hat H_\text{int}\,
    \left[1-\frac{\hat p^2}{2m^2c^2}-\frac{GM}{dc^2} \left(1-\frac{x}{d}+\frac{x^2}{d^2} \right)\right].
    \label{eq:hamiltonian_GR2}
\end{align}
We assume that the COM degrees of freedom is initially prepared in a coherent state at $t=0$:
\begin{align}
|\psi_\text{ini}\rangle=|\alpha\rangle~.
\end{align}
Here, the coherent state is an eigenstate of the annihilation operator $\hat a=\sqrt{\frac{m \Omega}{2\hbar}}\hat x+i \frac{\hat p}{\sqrt{2\hbar m \Omega}}$, where $\Omega$ is a reference frequency that determines the width of the initial Gaussian wavepacket. 

\begin{figure*}[htbp]
    \centering
    \includegraphics[width=1\textwidth]{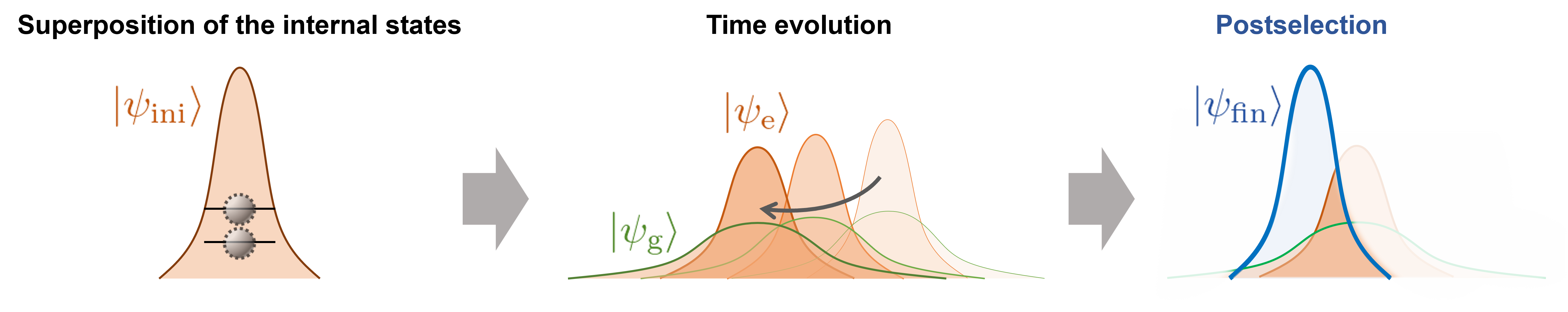}
    \caption{
    Schematic of Ramsey interferometry with COM postselection using coherent states. The clock's COM is initially prepared in $|\psi_{\ini}\rangle=|\alpha\rangle$ and evolves in a weak gravitational field, where the wavepacket spreads and shifts due to gravity. Postselection onto another coherent state $|\psi_{\rm fin}\rangle=|\beta\rangle$ can then produce an effective time advance, detected through the final Ramsey measurement.}
    \label{fig:coherent}
\end{figure*}

Let us now consider Ramsey interferometry with the coherent state quantum clock.
We first prepare the internal degrees of freedom in a superposition state, as described in Eq.~\eqref{eq:Ramsey2}, and then let the system evolve under the Hamiltonian in Eq.~\eqref{eq:hamiltonian_GR}, as given in Eq.~\eqref{eq:Ramsey3}.
As illustrated in Fig.~\ref{fig:coherent}, the wavepacket spreads due to free propagation, while its center moves toward the negative $x$ direction due to gravitational attraction.

We first examine how time delay arises in conventional Ramsey interferometry without postselection. In the following subsection, we will show how postselection can instead lead to an effective time advance.

The excitation probability $\mathcal{P}$, given in Eq.~\eqref{eq:Ramsey_prob}, is characterized by the Ramsey interference $\langle \psi_g |\psi_e\rangle$, which is now given by
\begin{align}
&\langle\psi_g(t)|\psi_e(t)\rangle
=
\frac{
e^{-i\omega t \Big(1-\frac{3\gamma}{4} \Big)}
}{
\sqrt{u_g^* u_e-v_g^*v_e}
}\nonumber\\
&\qquad\times 
\exp\Bigg[\left(\frac{1}{u_g^* u_e-v_g^*v_e}-1\right)
\Big|\alpha-\frac{\Delta}{2\sqrt{2}}\Big|^2
\nonumber\\
&\qquad\qquad
+i~\frac{\mathrm{Im}\bigg[(u_g v_e -v_g u_e)\Big(\alpha^*-\frac{\Delta}{2\sqrt{2}} \Big)^2\bigg]}{u_g^* u_e-v_g^*v_e}
\Bigg]
~.
\label{eq:Ramsey_interference_coherent}
\end{align}
Here, we introduced the gravitational coupling $\gamma$ and the distance parameter $\Delta$ as
\begin{align}
\gamma=\frac{GM}{c^2 d}, \quad \Delta=d \sqrt{\frac{m\Omega}{\hbar}}
~,
\end{align}
and also defined
\begin{align}
u_g
&=\cosh(\kappa t)
+\frac{i}{2}
\left(\frac{\Omega}{\kappa}-\frac{\kappa}{\Omega}\right)
\sinh(\kappa t),
\\
v_g
&=\frac{i}{2}
\left(\frac{\Omega}{\kappa}+\frac{\kappa}{\Omega}\right)
\sinh(\kappa t), 
\nonumber 
\\
u_e
&=\cosh\left(
\kappa \sqrt{1-\epsilon^2}\,t
\right)
\nonumber\\
&\quad+\frac{i}{2}
\left(
\frac{\Omega}
{\kappa}
\sqrt{\frac{1-\epsilon}
{1+\epsilon}}
-\frac{\kappa}{\Omega}
\sqrt{\frac{1+\epsilon}
{1-\epsilon}}
\right)
\sinh\left(
\kappa\sqrt{1-\epsilon^2}\, t
\right),
\\
v_e
&=\frac{i}{2}
\left(
\frac{\Omega}
{\kappa}
\sqrt{\frac{1-\epsilon}
{1+\epsilon}}
+\frac{\kappa}{\Omega}
\sqrt{\frac{1+\epsilon}
{1-\epsilon}}
\right)
\sinh\left(
\kappa \sqrt{1-\epsilon^2}\, t
\right),
\end{align}
with 
\begin{align}
\epsilon=\frac{\hbar \omega}{mc^2}, \quad \kappa=\sqrt{\frac{2GM}{d^3}}.
\end{align}
This interference term can be calculated exactly using the approximate quadratic Hamiltonian in Eq.~\eqref{eq:hamiltonian_GR2}.

\begin{figure}[htbp]
    \centering
    \includegraphics[width=1\linewidth]{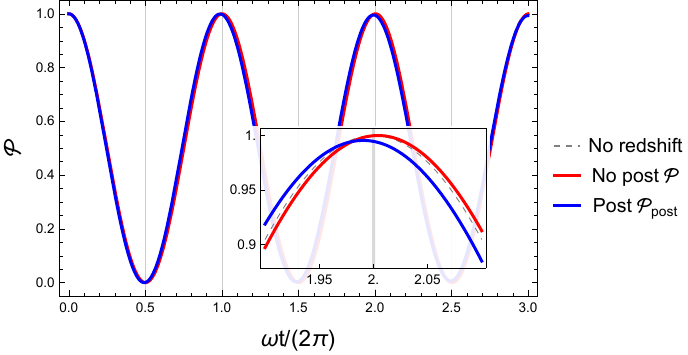}
    \caption{
    The measurement time dependence of the excitation probability 
    evaluated using Eq.~\eqref{eq:Ramsey_prob},~\eqref{eq:Ramsey_interference_coherent},~\eqref{eq:Ramsey_prob_PS},~\eqref{eq:interference_PS_GR_coherent}.
    The gray dashed, red, blue curve shows the result without the redshift effect, without the postselction and with the postselection. The parameters are given by $\epsilon=0.08,~\gamma=10^{-4},~\Omega/\omega=0.1,~\Delta=50,~\alpha=0$ and $\beta=3$.
    We can observe that the red curve oscillates slower than the gray dashed curve, representing the time dilation. On the other hand, the blue curve oscillates faster than the gray dashed curve, indicating the effective time advance.
    }
    \label{fig:prob_inicoherent_SRGR}
\end{figure}

Fig.~\ref{fig:prob_inicoherent_SRGR} shows the excitation probability as a function of the measurement time $t$. The gray dashed curve presents the case without redshift ($\epsilon,\gamma,\kappa\to 0$), while the red curve corresponds to conventional Ramsey interferometry without postselection.
The parameters are given by 
$\epsilon=0.08,~\gamma=10^{-4},~\Omega/\omega=0.1,~\Delta=50$ and $\alpha=0$. We observe that the red curve oscillates slightly more slowly than the gray dashed curve, indicating the time delay due to relativistic effects.

\begin{figure}[htbp]
    \centering
    \includegraphics[width=1\linewidth]{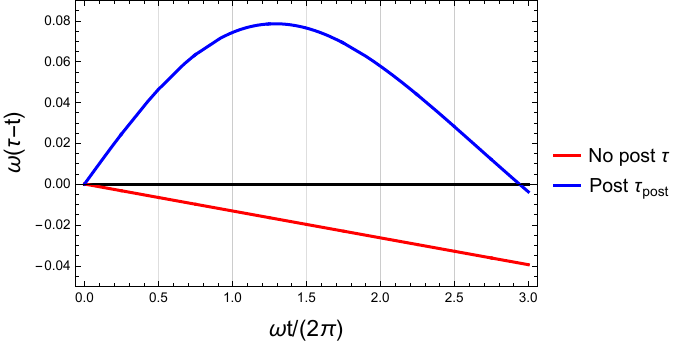}
    \caption{
    The difference between the proper time and measurement time, $\omega(\tau-t)$ evaluated using Eq.~\eqref{eq:tau_GR},~\eqref{eq:Ramsey_interference_coherent},~\eqref{eq:tau_PS_GR},~\eqref{eq:interference_PS_GR_coherent}.
    The red and blue curve shows the result without the postselction and with the postselection. The parameters are given by $\epsilon=0.08,~\gamma=10^{-4},~\Omega/\omega=0.1,~\Delta=50,~\alpha=0$ and $\beta=3$.
    We can observe that the red curve always takes the negative value, indicating the time delay. On the other hand, the blue curve takes the positive value at earlier time, indicating the effective time advance.
    }
    \label{fig:propertime_inicoherent_SRGR}
\end{figure}

The proper time can also be extracted from the phase of the Ramsey interference term, as given in Eq.~\eqref{eq:tau_GR}.
The red curve in Fig.~\ref{fig:propertime_inicoherent_SRGR} shows the difference between the proper time and the measurement time, expressed as $\omega(\tau-t)$, as a function of $\omega t/(2\pi)$, which is extracted from Ramsey interference without postselection.
The parameters are the same as those in Fig.~\ref{fig:prob_inicoherent_SRGR}: 
$\epsilon=0.08,~\gamma=10^{-4},~\Omega/\omega=0.1,~\Delta=50$ and $\alpha=0$.
A negative value of $\omega(\tau-t)$ indicates that the proper time $\tau$ is shorter than the reference time $t$, corresponding to a time delay. As shown in the figure, the red curve always remains negative, confirming the time delay.


\subsection{Coherent state time advance}

Next, we examine how an effective time advance arises in the Ramsey interferometry with postselection.
After the clock evolves under the Hamiltonian, we perform postselection on the COM state, as depicted in Fig.~\ref{fig:coherent}. We assume the postselected state is also a coherent state, given by
\begin{align}
    |\psi_\text{fin}\rangle=|\beta\rangle
    ~.
\end{align}
Then, the conditional excitation probability $\mathcal{P}^B_\pos$ is given from the ratio of inner products, $\frac{\langle \psi_\text{fin}|\psi_e \rangle}{\langle \psi_\text{fin}|\psi_g\rangle}$. 
The ratio can be now explicitly calculated as 
\begin{align}
&\frac{\langle \psi_\text{fin}|\psi_e \rangle}{\langle \psi_\text{fin}|\psi_g\rangle}
\nonumber
\\
&=\sqrt{\frac{u_g}{u_e}}\exp\Bigg[
-i\omega t\Big(1-\frac{3\gamma}{4}\Big)
\nonumber\\
&\quad+
\left(\frac{1}{u_e}-\frac{1}{u_g}\right)
\left(\alpha-\frac{\Delta}{2\sqrt{2}}\right)
\left(\beta^*-\frac{\Delta}{2\sqrt{2}}\right)
\nonumber\\
&\quad+
\frac{1}{2}
\left(\frac{v_e}{u_e}-\frac{v_g}{u_g}\right)
\left\{
\left(\alpha-\frac{\Delta}{2\sqrt{2}}\right)^2
+
\left(\beta^*-\frac{\Delta}{2\sqrt{2}}\right)^2
\right\}
\Bigg],
\label{eq:interference_PS_GR_coherent}
\end{align}
This is an exact analytical result within the approximate quadratic Hamiltonian~\eqref{eq:hamiltonian_GR2}.

The blue curve in Fig.~\ref{fig:prob_inicoherent_SRGR} shows 
the excitation probability with postselection
onto the coherent state with $\beta=3$, while the other parameters are the same as those specified above.
We observe that the blue curve oscillates slightly faster than the gray dashed curve, in contrast to the slower oscillation of the red curve. This demonstrates that postselection can induce an effective time advance of the quantum clock.

The effective proper time can also be extracted from the ratio $\frac{\langle \psi_\text{fin}|\psi_e \rangle}{\langle \psi_\text{fin}|\psi_g\rangle}$, as given in Eq.~\eqref{eq:tau_PS_GR}.
Moreover, to first order in the Dyson expansion, the effective proper time is approximately given by
\begin{align}
    \tau_\pos^B
    &\simeq
    t\Bigg(
    1
    -\frac{\epsilon\Omega}{4\omega}\,
    \mathrm{Re}\bigg[
    \frac{1+i\Omega t/2-(\alpha-\beta^*)^2}{(1+i\Omega t/2)^2}
    \bigg]\nonumber\\
    &\hspace{20mm}-
    \gamma\,
    \mathrm{Re}\bigg[
    1-\frac{\alpha+\beta^*}{\sqrt{2}\Delta}
    \bigg]
    \Bigg)
    ~.
\end{align}
To obtain this expression, we evaluated the weak values of the interaction-picture operators in Eq.~\eqref{eq:tau_GR_aprx}, performed the time integration, and neglected higher-order terms in $1/\Delta$.
The second and third terms represent the special relativistic and gravitational contributions, respectively, arising from $(\hat p_I^2)_{\widetilde w}/(2m^2c^2)$ and $-(\hat\Phi_I)_{\widetilde w}/c^2$.

Furthermore, let us set $\alpha=0$ and examine the conditions for effective time advance. In the early time limit $\Omega t\ll 1$, the real part of the special relativistic weak value is given by 
$\mathrm{Re}[(\hat p_I^2)_{\widetilde w}]/(2m^2c^2)=\epsilon\Omega(1-\mathrm{Re}[\beta^2])/(4\omega)$
which becomes negative when $\mathrm{Re}[\beta^2]>1$. 
In contrast, in the late time $\Omega t\gg 1$, we obtain $\mathrm{Re}[(\hat p_I^2)_{\widetilde w}]/(2m^2c^2)=
\epsilon\Omega(1+\mathrm{Re}[\beta^2])/[\omega(\Omega t)^2]
$, which becomes negative when $\mathrm{Re}[\beta^2]<-1$, although its magnitude is suppressed as $(\Omega t)^{-2}$. Thus, the conditions for special relativistic time advance in the early and late times do not overlap, making it impossible to satisfy the time advance condition in both time regimes.
On the other hand, the condition for GR time advance reduces to $\mathrm{Re}[\beta]>\sqrt{2}\Delta\gg 1$, requiring postselection onto a COM state far from its initial position. Such postselection is highly unlikely to succeed in the present setup, since our calculation assumes $x/d\ll1$.
Therefore, the most promising approach to observing effective time advance is to choose $\mathrm{Re}[\beta^2]>1$ and focus on the special-relativistic contribution in the early time regime.

Finally, the blue curve in Fig.~\ref{fig:propertime_inicoherent_SRGR} shows the time dependence of $\omega(\tau_\pos-t)$. 
The parameters are the same as before, while the coherent parameter of the postselected state is set to $\beta=3$.
We observe that the blue curve is positive at early times, $\omega(\tau_\pos-t)>0$, indicating that the effective proper time exceeds the reference time $t$. This demonstrates the effective time advance induced by postselection.
At later times, however, the blue curve becomes negative, indicating that the clock undergoes time dilation even after postselection.
This behavior is consistent with our previous discussion: when $\mathrm{Re}[\beta^2]>1$, special-relativistic time advance can occur at early times but not at late times.

In this section, we have demonstrated that effective time advance can arise even for simple coherent states, without using Schrödinger-cat-like superpositions between $|0\rangle$ and $|p\rangle$. This can be understood intuitively by noting that even a single Gaussian wavepacket contains a range of momentum components. During free evolution, these components propagate differently, and postselection onto a spatially displaced Gaussian wavepacket changes how they interfere, rather than simply selecting 
a momentum-space cat state $(|0\rangle-|p\rangle)/\sqrt{2}$. 
For a sufficiently large displacement, this interference produces a negative weak value of the squared momentum, just as in Sec.~III, where two distinct momentum branches were used. Thus, the coherent-state example shows that the same mechanism of quantum interference and postselection can produce effective time advance.

\section{VI. Time advance in a hydrogen-like atom}



In the previous sections, we demonstrated an effective time advance by following the standard Ramsey interferometry protocol procedurally.
We now turn to the underlying physical picture and examine how this effect manifests in the dynamics of the quantum clock. 
To this end, we consider a hydrogen-like atom as a concrete physical model and examine the wavefunction of its internal degrees of freedom and the associated electric dipole moment.
We show that the postselected internal state is a superposition of wavefunctions that have undergone different relativistic time dilations. The quantum interference between these states gives rise to an effective time advance, which is also manifested in the oscillation of the atom's electric dipole moment.

The atom consists of two particles with charges $q$ and $-q$. 
Its motion can be decomposed into the COM motion and the relative motion of two particles, the latter representing the internal degrees of freedom of the quantum clock.
The wavefunctions for the relative motion are characterized by the usual quantum numbers $(n, \ell, m)$. We identify the ground state $|g\rangle$ and the excited $|e\rangle$ with the $1s$- and $2p_z$-orbital states, corresponding to $(n, \ell, m)=(1,0,0)$ and $(2,1,0)$, respectively.
The corresponding wavefunctions, $\chi_g $ and $\chi_e$
are given by
\begin{align}
\chi_g (\bm{r})&=\frac{1}{\sqrt{\pi a_0^3}} e^{-r/a_0},  
\label{eq:wavefunction_g}
\\
\chi_e (\bm{r})&=\frac{1}{4\sqrt{2\pi a_0^3}} \frac{z}{a_0} e^{-r/(2a_0)},
\label{eq:wavefunction_e}
\end{align}
where $r$ is the magnitude of the relative position $\bm{r}=(x,y,z)$, and
$a_0$ is the Bohr radius of the hydrogen-like atom.  

\begin{figure}
    \centering
    \includegraphics[width=0.9\linewidth]{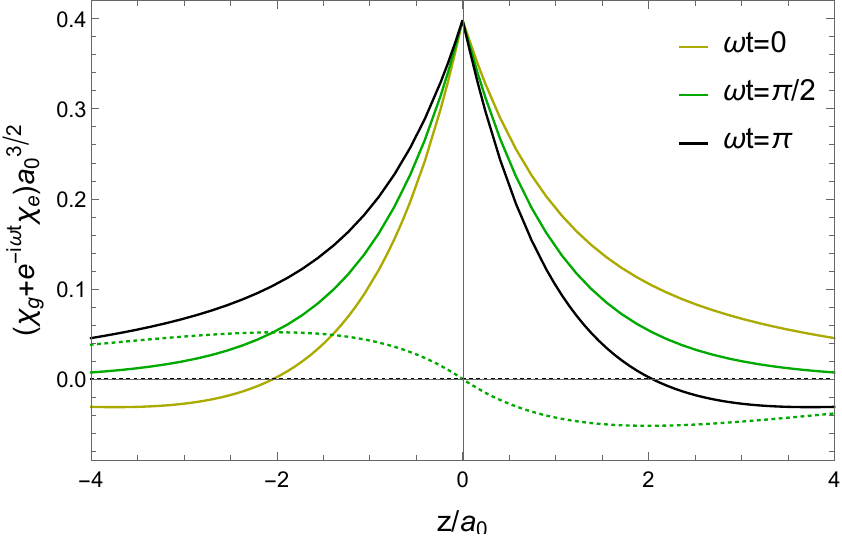}
    \caption{The time evolution of the superposed internal state given in Eq.~\eqref{eq:+state_evolve}. The horizontal axis shows $z/a_0$, while we set $x=y=0$. The solid and dotted line shows the real and imaginary parts. }
    \label{fig:atom_catstate_evolution}
\end{figure}

Let us first examine how the evolution of the internal clock can be described in this atomic model. In Ramsey interferometry, we first prepare the internal degrees of freedom in the superposition state $(|g\rangle+|e\rangle)/\sqrt{2}$, as described in Eq.~\eqref{eq:Ramsey2}.
For later use, we first consider how this superposition evolves under the internal Hamiltonian $\hat H_{\rm int}$. Its time evolution is given by
\begin{align}
    e^{-i\hat H_{\rm int}t/\hbar}
    \frac{|g\rangle+|e\rangle}{\sqrt{2}}
    \sim
    \frac{|g\rangle+e^{-i\omega t}|e\rangle}{\sqrt{2}}
    ~,
    \label{eq:+state_evolve}
\end{align}
where we have neglected an overall phase.
In the atomic wavefunction picture, the ground state $\chi_g$ and excited state $\chi_e$ have even and odd parity with respect to the $z$ direction, respectively. According to the above expression, the relative phase between these two states changes during the time evolution, causing the spatial probability distribution to oscillate along the $z$ direction.
Fig.~\ref{fig:atom_catstate_evolution} shows the time evolution of the rescaled wavefunction given in Eq.~\eqref{eq:+state_evolve}. The horizontal axis represents $z/a_0$, with $x=y=0$. The solid and dotted lines represent the real and imaginary parts of the wavefunction, respectively. Different colors correspond to different times, $\omega t=0,\pi/2$, and $\pi$.
We observe that the wavefunction is initially biased toward the positive $z$ direction, while its distribution gradually shifts toward the negative $z$ direction as time evolves. In this sense, the atom acts as a clock, with its internal dynamics manifested as periodic oscillations of the spatial probability distribution along the $z$ direction.

Let us also examine the electric dipole moment of this atomic model to characterize the spatial asymmetry of the wavefunction. The electric dipole moment has only a nonzero $z$ component, given by
\begin{align}
d(t)&=-q\langle\hat z\rangle
=-q\langle g|\hat z|e\rangle\cos(\omega t)
\nonumber
\\&=-\frac{128\sqrt{2}}{243}qa_0\cos(\omega t),
\label{eq:dipole}
\end{align}
where the numerical factor is calculated from the orbital wavefunctions given above.
Thus, the oscillation of the electric dipole moment directly reflects the evolution of the internal clock and provides a concrete observable through which to illustrate time dilation and effective time advance.

Now, let us consider the Ramsey interferometry with the postselection.
We assume that the COM motion and the internal relative motion are initialy in 
\begin{align}
    |\Psi_\ini\rangle=|g\rangle \left(\sqrt{\lambda} |0\rangle+\sqrt{1-\lambda} |p\rangle\right)
    ~,
\end{align}
where $0 \leq \lambda \leq 1$. 
Here, the COM motional states $|0\rangle$ and $|p\rangle$ have
average momenta of zero and $p$ 
, respectively, and are assumed to be sufficiently orthogonal to each other. 
We then perform the Ramsey interferometry followed by postselection on the COM state. We first prepare the superposition of the internal state as $(|g\rangle+|e\rangle)/\sqrt{2}$ 
and let the system evolve in time. We then postselect the COM state onto
\begin{align}
    |\psi_\text{fin} \rangle=\frac{1}{\sqrt{2}}(|0\rangle-|p \rangle)
    ~.
\end{align}
In the following, for simplicity, we focus only on the special relativistic time dilation.
The relative motional state is then projected onto the following wavefunction.
\begin{align}
\chi_\pos
&= \frac{1}{N}\Bigg[\sqrt{\lambda} \Big(\frac{\chi_g +e^{-i\omega t} \chi_e}{\sqrt{2}} \Big)
\nonumber
\\
&\quad 
-\sqrt{1-\lambda} \, e^{-i p^2 t/(2\hbar m)} \Big(\frac{\chi_g +e^{-i\omega \tau} \chi_e}{\sqrt{2}} \Big)\Bigg],
\end{align}
where 
$N$ is the normalization, and
\begin{align}
\tau = t \left(1-\frac{p^2}{2m^2 c^2} \right).
\end{align}
We focus on the time scale $t$ such that $p^2t/(2\hbar m)\ll1$ and $p^2 \omega t/(2m^2 c^2) \ll 1$. 
The postselected wavefunction then reduces to
\begin{align}
\chi_\pos
&\approx \frac{1}{N}\Big[\sqrt{\lambda} \,\chi_\text{no delay}
-\sqrt{1-\lambda} \,\chi_\text{delay}
\Big]~,
\label{eq:post_app}
\end{align}
where 
$N$ is chosen as
\begin{align}
N
=
\sqrt{
1
-\sqrt{\lambda(1-\lambda)}
\left[
1+
\cos\!\left(\omega(t-\tau)\right)
\right]
},
\end{align}
and we define
\begin{align}
    \chi_\text{no delay} =\frac{\chi_g +e^{-i\omega t} \chi_e}{\sqrt{2}},
    \quad
    \chi_\text{delay}=\frac{\chi_g +e^{-i\omega \tau} \chi_e}{\sqrt{2}}~.
\end{align}
The derivation of Eq.~\eqref{eq:post_app} is presented in Appendix.
Comparing these expressions with Eq.~\eqref{eq:+state_evolve}, we see that the wavefunctions $\chi_\text{no delay}$ and $\chi_\text{delay}$ oscillate according to the reference time $t$ and the relativistic proper time $\tau$, respectively. 
The postselected internal state $\chi_\pos$ is thus given by a superposition of two wavefunctions, one evolving without time delay, $\chi_\text{no delay}$, and the other experiencing relativistic time delay, $\chi_\text{delay}$.

\begin{figure}[htbp]
    \centering
    \includegraphics[width=1.0\linewidth]{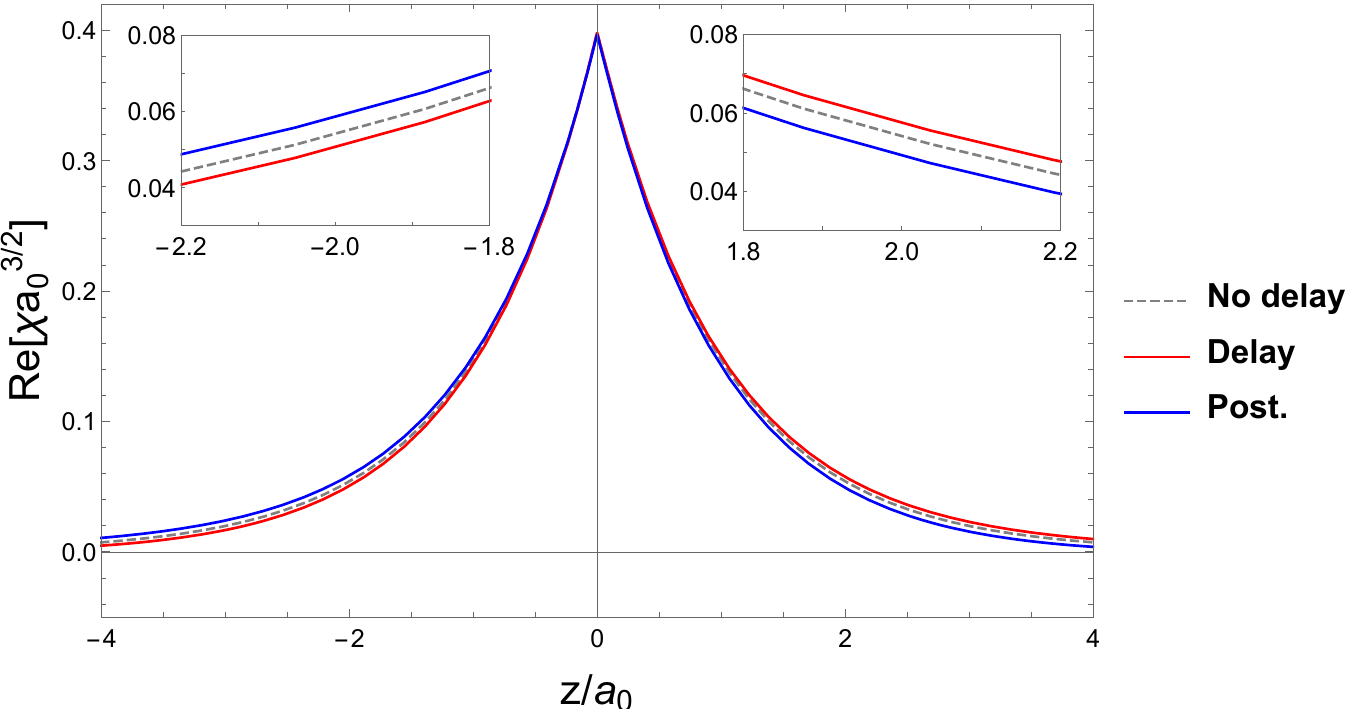}
    \includegraphics[width=1.0\linewidth]{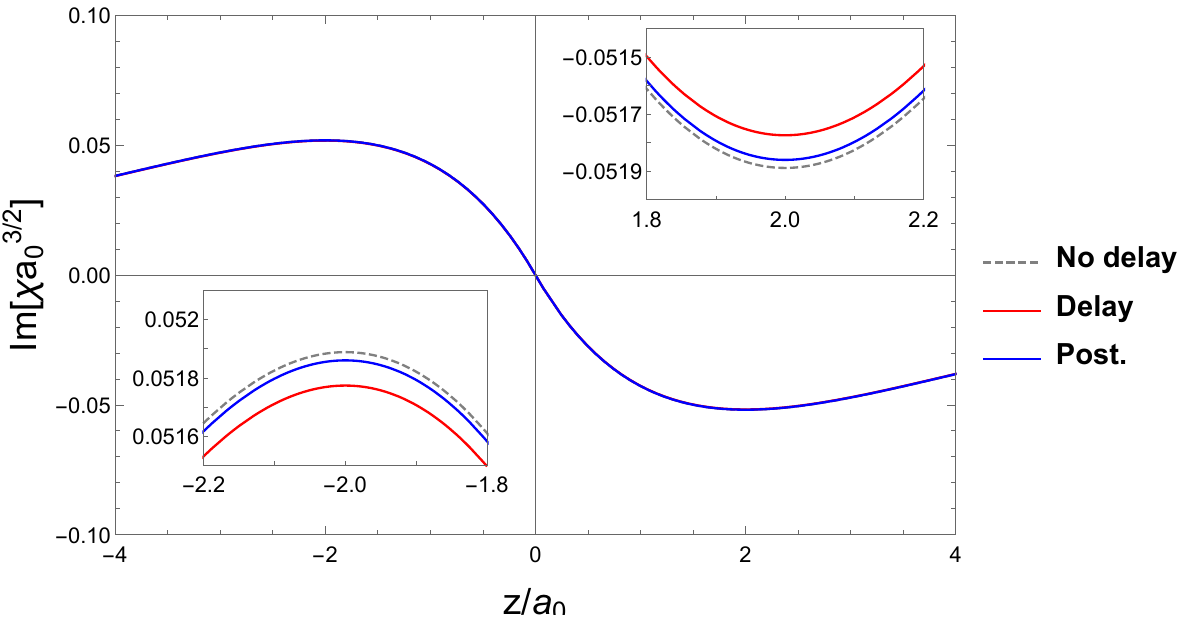}
    \caption{The upper panel and the lower panel show the real parts and the imaginary parts of the wavefunctions $\chi_\text{no delay}$, 
    $\chi_\text{delay}$ and $\chi_\text{post}$ as a function of $z$, respectively. 
    The insets in each panel show magnified views of the behavior near $z/a_0 = \pm 2.0$.
    The parameters are $\lambda=0.75$, $p^2/(2m^2 c^2)=0.0047$ and $\omega t=9\pi/2$.}
    \label{fig:wave}
\end{figure}

Fig.~\ref{fig:wave} compares the wavefunctions $\chi_\text{no delay}$, $\chi_\text{delay}$ and 
$\chi_\pos$ 
at $x=y=0$, represented by the dashed gray, red and blue curves, respectively. The parameters are chosen as $\lambda=0.75$, $p^2/(2m^2 c^2)=0.0047$ and $\omega t=9\pi/2$.
While the imaginary parts of the three wavefunctions are nearly identical, their real parts exhibit noticeable differences.
At this time, $\text{Re}[\chi_\text{no delay}]$ is almost symmetric about $z=0$, whereas $\text{Re}[\chi_\text{delay}]$ is biased towards positive $z$ 
and $\text{Re}[\chi_\pos]$ towards negative $z$. 
Since their imaginary parts are nearly identical, the differences in their real parts lead to corresponding differences in the probability distributions.

These opposite spatial shifts in Fig.~\ref{fig:wave} provide an intuitive picture of time delay and time advance. As discussed in Eq.~\eqref{eq:+state_evolve}, interference between $\chi_g$ and $\chi_e$ causes the atom's probability distribution to oscillate along the $z$ directions. Fig.~\ref{fig:wave} captures a moment when the probability distribution of $\chi_{\text{no delay}}$ is shifting from positive to negative $z$. The distribution of $\chi_{\text{delay}}$ still favors positive $z$, indicating that its oscillation lags behind the reference clock $\chi_{\text{no delay}}$. In contrast, $\chi_\pos$ is already biased toward negative $z$, indicating that its oscillation is ahead of $\chi_{\text{no delay}}$. Remarkably, this effective time advance emerges from the destructive interference between $\chi_{\text{no delay}}$ and $\chi_{\text{delay}}$, even though neither wavefunction individually exhibits a time advance.

\begin{figure}[htbp]
    \centering
    \includegraphics[width=1.0\linewidth]{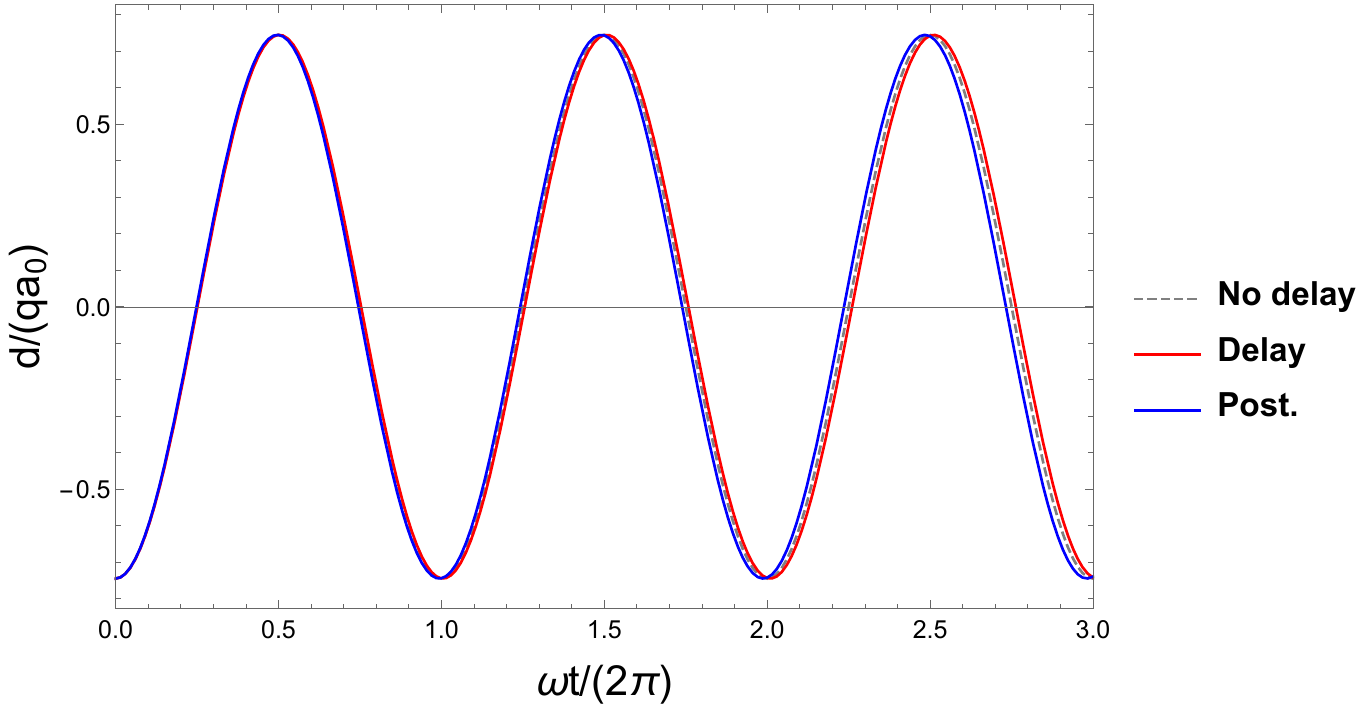}
    \caption{The temporal behaviors of $d=-q\langle\hat{z}\rangle$ for the wavefunctions $\chi_\text{no delay}$ (grey dashed), $ \chi_\text{delay}$ (red) and $\chi_\pos$ (blue). The parameters are $\lambda=0.75$ and $p^2/(2m^2 c^2)=0.0047$.}
    \label{fig:dipole}
\end{figure}

Fig.~\ref{fig:dipole}  directly shows the effective time advance in the oscillation of the atom's electric dipole moment. The figure presents the time evolution of $d(t)$, defined in Eq.~\eqref{eq:dipole}, in units of $qa_0$ for the three internal wavefunctions $\chi_{\text{no delay}}$, $\chi_{\text{delay}}$, and $\chi_\pos$. Compared with the reference oscillation of $\chi_{\text{no delay}}$, the dipole moment of $\chi_{\text{delay}}$ exhibits a phase delay, whereas that of $\chi_\pos$ exhibits a phase advance. Thus, the effective time advance is manifested not only in the phase of Ramsey interference but also in the dynamics of a concrete physical observable of the atom.


The hydrogen-like atom thus provides a concrete physical picture of effective time advance: although neither $\chi_{\text
{no delay}}$ nor $\chi_{\text
{delay}}$ exhibits a time advance, their superposition under postselection produces an advanced oscillation of the electric dipole moment. While we have focused on special relativistic effects in this section, the same mechanism is expected to operate when gravitational time dilation is included.

\section{Conclusion}

In this paper, we have shown that postselection of the COM state of a quantum clock can lead to an effective time advance, in contrast to the usual relativistic time dilation. Although each individual motional branch experiences an ordinary time delay, quantum interference between different branches can produce an effective time advance after postselection. We demonstrated this effect for both special relativistic and gravitational time dilation and showed that it can be understood in terms of anomalous weak values.

We further demonstrated that effective time advance can arise even when both the initial and postselected COM states are simple coherent states, without requiring specially prepared Schrödinger-cat states. By solving the time evolution exactly within an approximate Hamiltonian, we showed that the effect persists when the COM dynamics is explicitly taken into account. We also considered a hydrogen-like atom as a concrete model of a quantum clock and illustrated how quantum interference between internal wavefunctions gives rise to the effective time advance, which is reflected in the oscillation of the electric dipole moment.

An important direction for future work is the experimental verification of this effect. In particular, the special relativistic time advance could, in principle, be tested by combining Ramsey interferometry with postselection of the COM state of an atomic clock. 
A key challenge is to achieve a sufficiently high postselection success probability while maintaining enough interference contrast to resolve the anomalous phase shift. Furthermore, our analysis relies on leading-order relativistic corrections and, in the gravitational case, a quadratic approximation to the potential. Assessing the effects of higher-order corrections and experimental imperfections will be important for developing a realistic experimental proposal. Such experiments could provide a new perspective on how relativistic time dilation manifests itself in quantum systems.

Note added. During the preparation of this manuscript, we became aware of related work by Saldanha and Yokomizo \cite{Saldanha2026}, which also demonstrates that postselection can lead to anomalous effective proper times in a quantum clock.

\begin{acknowledgments}
\textit{Acknowledgments.---}
We would like to thank Kota Fujii and Yuka Shiomatsu for fruitful discussions.
Y.K. acknowledges support from JSPS KAKENHI Grant No. JP26K17151 and the JSPS Overseas Research Fellowship.
A.M. acknowledges support from JSPS KAKENHI Grant Nos. JP23K13103 and JP23H01175.
T.F. acknowledges support from JSPS KAKENHI Grant Nos. JP26H00407 and JP23K03424.
\end{acknowledgments}

\begin{appendix}

\section{Appendix}
\section{Derivation of the approximate $\chi_\pos$}
\label{app:post_app}

Here, we derive the approximated form of $\chi_\pos$ given in Eq. \eqref{eq:post_app} for a short time $t$ such that  $p^2t/(2m\hbar) \ll 1$ and $p^2 \omega t/(2m^2c^2) \ll 1$. 
We can evaluate the postselected wavefunction $\chi_\pos$ as 
\begin{widetext}
\begin{align}
\chi_\pos
&= \frac{1}{N}\Bigg[\sqrt{\lambda} \Big(\frac{\chi_g +e^{-i\omega t} \chi_e}{\sqrt{2}} \Big)
-\sqrt{1-\lambda} \, e^{-i p^2 t/(2\hbar m)} \Big(\frac{\chi_g +e^{-i\omega \tau} \chi_e}{\sqrt{2}} \Big)\Bigg]
\nonumber
\\
& =\frac{1}{N}
\Bigg[\frac{\sqrt{\lambda}-\sqrt{1-\lambda} e^{-ip^2 t/(2\hbar m)} }{\sqrt{\lambda}-\sqrt{1-\lambda}}\left\{\sqrt{\lambda} \Big(\frac{\chi_g +e^{-i\omega t} \chi_e}{\sqrt{2}} \Big)-\sqrt{1-\lambda} \Big(\frac{\chi_g +e^{-i\omega \tau} \chi_e}{\sqrt{2}} \Big)\, \right\}
\nonumber
\\
&\quad
-\frac{\sqrt{\lambda(1-\lambda)} }{\sqrt{\lambda}-\sqrt{1-\lambda}}\Big(1-e^{-ip^2 t/(2\hbar m)} \Big)\Big(\frac{\chi_g+e^{-i\omega t} \chi_e}{\sqrt{2}} -\frac{\chi_g+e^{-i\omega \tau} \chi_e}{\sqrt{2}} \Big) \Bigg]
\nonumber
\\
& =\frac{1}{N}
\Bigg[\frac{\sqrt{\lambda}-\sqrt{1-\lambda} e^{-ip^2 t/(2\hbar m)} }{\sqrt{\lambda}-\sqrt{1-\lambda}}\Big(\sqrt{\lambda} \chi_\text{no delay}-\sqrt{1-\lambda} \chi_\text{delay} \Big)
\nonumber
\\
&\quad
- \frac{\sqrt{\lambda(1-\lambda)} }{\sqrt{\lambda}-\sqrt{1-\lambda}}\Big(1-e^{-ip^2 t/(2\hbar m)} \Big)\Big(1-e^{ip^2\omega t/(2m^2 c^2)} \Big)\frac{e^{-i\omega t} \chi_e}{\sqrt{2}} \Bigg]
\nonumber 
\\
& \simeq \frac{1}{N} \left(1+\frac{i\sqrt{1-\lambda}}{\sqrt{\lambda}-\sqrt{1-\lambda}} \frac{p^2 t}{2\hbar m} \right)
\Big(\sqrt{\lambda} \chi_\text{no delay}-\sqrt{1-\lambda} \chi_\text{delay}\Big)
\nonumber
\\
& \simeq \frac{1}{N} \exp\left[\frac{i\sqrt{1-\lambda}}{\sqrt{\lambda}-\sqrt{1-\lambda}} \frac{p^2 t}{2\hbar m} \right]
\Big(\sqrt{\lambda} \chi_\text{no delay}-\sqrt{1-\lambda} \chi_\text{delay}\Big).
\end{align}
\end{widetext}
In the fourth equality, we took the first order of $\frac{p^2 t}{2m\hbar}$ and $\frac{p^2 \omega t}{2m^2 c^2}$ by expanding the phase factors as
\begin{align}
\exp \Big[i\frac{p^2 t}{2m\hbar} \Big] \simeq 1+i\frac{p^2 t}{2m\hbar},
\quad 
\exp \Big[i\frac{p^2 \omega t}{2m^2c^2}\Big] \simeq 1+i\frac{p^2 \omega t}{2m^2 c^2}. 
\end{align}
In the fifth equality, we perfomed the approximation 
\begin{align}
1+\frac{i\sqrt{1-\lambda}}{\sqrt{\lambda}-\sqrt{1-\lambda}} \frac{p^2 t}{2\hbar m} \simeq \exp \Big[\frac{i\sqrt{1-\lambda}}{\sqrt{\lambda}-\sqrt{1-\lambda}} \frac{p^2 t}{2\hbar m}\Big].
\end{align}
Neglecting the total phase and adjusting the normalization, we get Eq. \eqref{eq:post_app}. 
\end{appendix}

\bibliography{ref.bib}

\end{document}